\documentclass[]{spie}  
\usepackage{appendix}
\usepackage[]{graphicx}
\usepackage{float}
\usepackage{multirow}
\usepackage[table,xcdraw]{xcolor}
\usepackage{mathtools}
\graphicspath{ {./figures/} }

\newcommand{\tabcell}[1]{\begin{tabular}{@{}c@{}}#1\end{tabular}}

\newcommand{\arcsec}{''}

\title{Liger at Keck Observatory: Design of the Data Reduction System and Software Interfaces} 

\author{
Nils Rundquist\supit{a,b},
Andrea Zonca\supit{f},
Arun Surya\supit{c},
Shelley A. Wright\supit{a,b}, 
Aaron Brown\supit{a,b},
Maren Cosens\supit{a,b},
Michael Fitzgerald\supit{d},
Chris Johnson\supit{d}, 
Marc Kassis\supit{g},
Renate Kupke\supit{h},
Kyle Lanclos\supit{g},
James E. Larkin\supit{d}, 
Kenneth Magnone\supit{d},
Rosalie McGurk\supit{g},
Ji Man Sohn\supit{d},
Gregory Walth\supit{e},
James Wiley\supit{a,b},
Sherry Yeh\supit{g}
\skiplinehalf
\supit{a} Center for Astrophysics \& Space Sciences, University of California San Diego, USA; \\
\supit{b} Department of Physics, University of California San Diego, USA; \\
\supit{c} Tata Institute of Fundamental Research, Mumbai, India \\
\supit{d} Physics \& Astronomy Department, University of California Los Angeles, CA 90095 USA; \\
\supit{e} Carnegie Observatories, 813 Santa Barbara St, Pasadena, CA 91101 USA \\
\supit{f} San Diego Supercomputer Center, University of California San Diego, USA; 
\\
\supit{g} W.M. Keck Observatory, 65-1120 Mamalahoa Hwy, Waimea, HI, USA;
\\
\supit{h} Department of Astronomy and Astrophysics, University of California Santa Cruz, USA;
}

\authorinfo{Further author information: (Send correspondence to N.R.)\\N.R...: E-mail: nrundqui@physics.ucsd.edu}

\begin{document} 
  \maketitle 

\begin{abstract}

Liger is a second generation near-infrared imager and integral field spectrograph (IFS) for the W. M. Keck Observatory that will utilize the capabilities of the Keck All-sky Precision Adaptive-optics (KAPA) system. Liger operates at a wavelength range of 0.81 $\mu$m - 2.45 $\mu$m and utilizes a slicer and a lenslet array IFS with varying spatial plate scales and fields of view resulting in hundreds of modes available to the astronomer.  Because of the high level of complexity in the raw data formats for the slicer and lenslet IFS modes, Liger must be designed in conjunction with a Data Reduction System (DRS) which will reduce data from the instrument in real-time and deliver science-ready data products to the observer. The DRS will reduce raw imager and IFS frames from the readout system and provide 2D and 3D data products via custom quick-look visualization tools suited to the presentation of IFS data. The DRS will provide the reduced data to the Keck Observatory Archive (KOA) and will be available to astronomers for offline post-processing of observer data.  We present an initial design for the DRS and define the interfaces between observatory and instrument software systems.

\end{abstract}




\section{Introduction}
\label{sec:intro}

The largest of all cat species is a hybrid animal of lion and a tiger, and so named for its similarly hybrid design inheriting aspects of predecessor instruments InfraRed Imaging Spectrograph (IRIS)\cite{Larkin1, Larkin2, Larkin3, Larkin4} and OH-Suppressing InfraRed Imaging Spectrograph (OSIRIS), Liger\cite{Wright2020,Wright2022} is a second generation near-infrared imager and integral field spectrograph (IFS) for W. M. Keck Observatory (WMKO).  Liger is designed to provide exceptional utility in accessing the new observational opportunities offered by the new Keck All-sky Precision Adaptive-optics (KAPA)\cite{Wizinowich} system.  Liger operates in the 0.81 $\mu$m - 2.45 $\mu$m wavelength range, and utilizes slicer and lenslet array IFS designs to offer various plate scales and fields of view for the spectrograph in addition to its imaging camera\cite{Cosens2020, Cosens2022, Whiley, Wright2022}.  The Liger imager offers a 20.5\arcsec x 20.5\arcsec Field of View (FoV) at a plate scale of 10 miliarcseconds (mas), and is designed to operate concurrently with the IFS.  The IFS mode plate scales range from 15 mas to 150 mas with spectral resolution options ranging from 4,000 to 10,000 and a large field of view range.  The operational modes are shown in Table \ref{tab:liger_modes}.

\begin{table}[]
\begin{center}
\caption{Operational modes overview of Liger. \label{tab:liger_modes}}
\begin{tabular}{|cccccc|}
\hline

\multicolumn{2}{|c|}{\textbf{Mode}} & \multicolumn{1}{c|}{\textbf{\tabcell{Spatial\\Sampling}}} & \multicolumn{1}{c|}{\textbf{\tabcell{Field\\of View}}} & \multicolumn{1}{c|}{\textbf{\tabcell{Spectral\\Resolution (R)}}} & \textbf{\tabcell{Wavelength\\Bandpass}} \\
\hline
\multicolumn{2}{|c|}{\textbf{Imager}} & \multicolumn{1}{c|}{10 mas} & \multicolumn{1}{c|}{20.5\arcsec x20.5\arcsec} & \multicolumn{1}{c|}{Set By Filter} & 5\%, 20\%, 40\% \\
\hline
\multicolumn{2}{|c}{\textbf{Lenslet IFS}} &  &  &  &  \\
\hline
\multirow{2}{*}{\tabcell{Spatial\\Elements}} & \multicolumn{1}{|c|}{128 x 128} & \multicolumn{1}{c|}{\tabcell{15 mas \\ 31 mas}} & \multicolumn{1}{c|}{\tabcell{1.9\arcsec x1.9\arcsec \\ 3.9\arcsec x3.9\arcsec}} & \multicolumn{1}{c|}{4000} & 5\% \\
\cline{2-6}
& \multicolumn{1}{|c|}{16 x 128} & \multicolumn{1}{c|}{\tabcell{15 mas \\ 31 mas}} & \multicolumn{1}{c|}{\tabcell{0.2\arcsec x1.9\arcsec \\ 0.5\arcsec x3.9\arcsec}}  & \multicolumn{1}{c|}{4000, 8000, 10000} & 5\%, 20\%, 40\% \\
\hline
\multicolumn{2}{|c}{\textbf{Slicer IFS}} &  &  &  &  \\
\hline
\multirow{2}{*}{\tabcell{Spatial\\Elements}} & \multicolumn{1}{|c|}{88 x 45} & \multicolumn{1}{c|}{\tabcell{75 mas \\ 150 mas}} & \multicolumn{1}{c|}{\tabcell{6.6\arcsec x3.4\arcsec \\ 13.2\arcsec x6.8\arcsec}} & \multicolumn{1}{c|}{4000} & 5\%, 20\% \\
\cline{2-6}
& \multicolumn{1}{|c|}{44 x 45} & \multicolumn{1}{c|}{\tabcell{75 mas \\ 150 mas}} & \multicolumn{1}{c|}{\tabcell{3.3\arcsec x3.4\arcsec \\ 6.6\arcsec x6.8\arcsec}} & \multicolumn{1}{c|}{4000, 8000, 10000} & 5\%, 20\%, 40\% \\
\hline
\end{tabular}
\end{center}
\end{table}

\subsection{Motivation for the Data Reduction System}

\begin{figure}[h]
    \centering
    \includegraphics[scale=0.5]{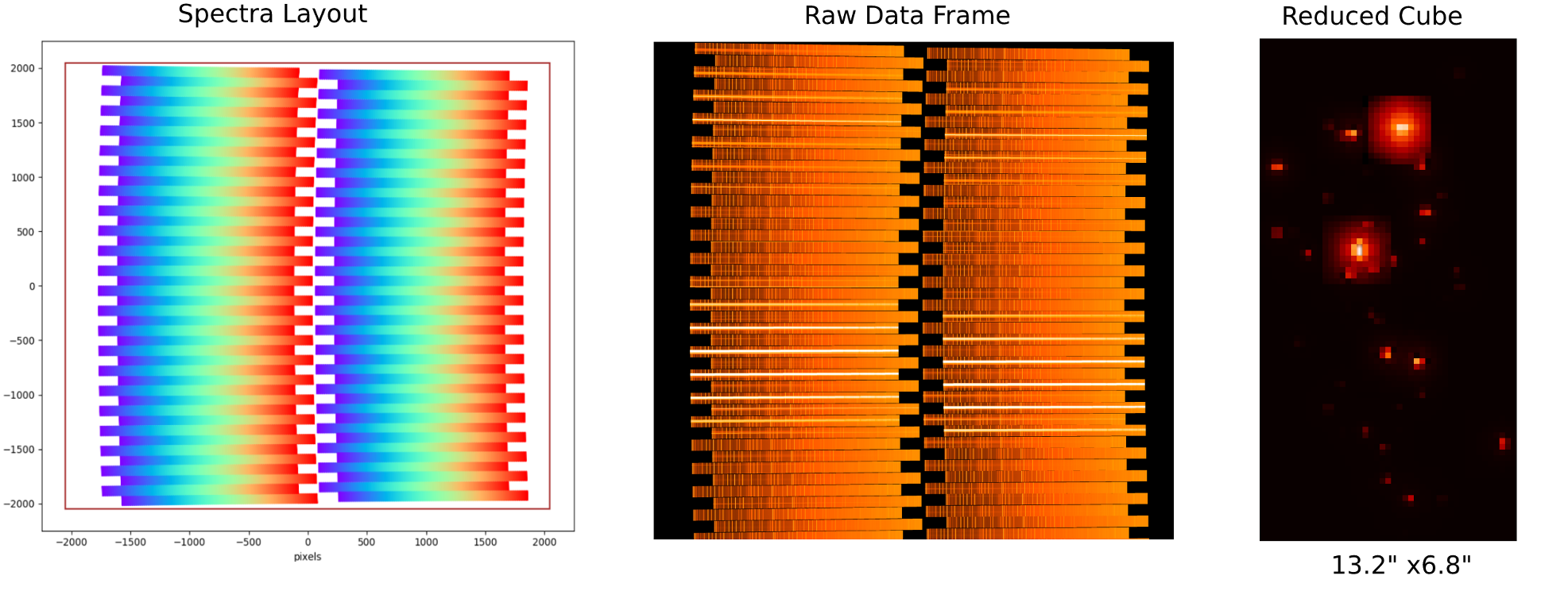}
    \includegraphics[scale=0.25]{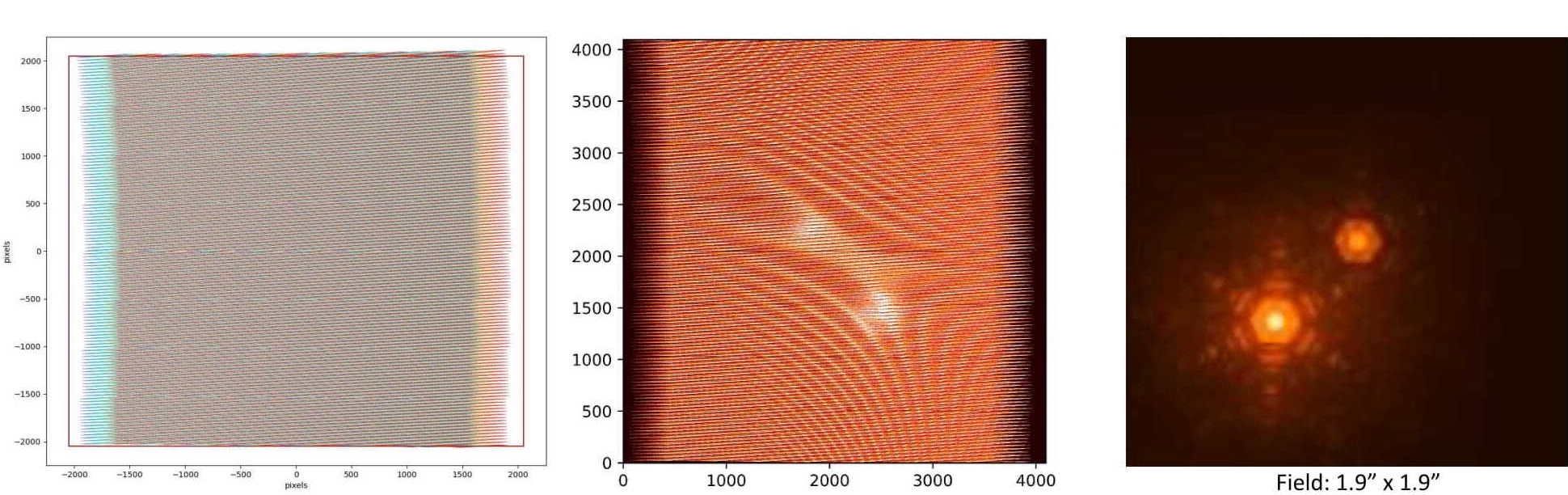}
    \caption{Simulations of the Liger Slicer (above) and Liger Lenslet (below) spectral distribution across the IFS detector (Left, Middle) and a 2D image slice of the reduced data cube following the reduction of the raw detector data (Right) using the 150mas mode for the slicer, and the 15 mas mode for the lenslet in the Kbb filter. In order to accurately present IFS data products raw detector frames must be processed and reduced by a robust data reduction software pipeline. The simulated slicer field consists of source brightness down to 23rd magnitude (Vega) over 20 sources, and the lenslet field consists of two sources. Liger performance was simulated with an integration time of 1 hour and 15 minutes. \label{fig:slicer_fig}}
\end{figure}

Liger is an important development in instruments for observational astronomy; through offering unique science capabilities at high spectral and spatial resolution many novel avenues of research are made available\cite{Wright2022}.  Liger's complex optical design\cite{Cosens2020, Cosens2022, Whiley, Wright2022} demands the need for a robust Data Reduction System (DRS) in order to provide usable science data products for Liger's wide variety of operational modes. The complexity of IFS data products in particular necessitates advanced processing of the raw detector data into three-dimensional (X, Y, $\lambda$) data cubes usable by astronomers.  Exemplifying this, we generate simulated raw data frames for the Liger slicer and lenslet IFS modes for testing and development by utilizing and advancing from simulation code used for IRIS and OSIRIS instruments\cite{Wright1, Wright2, Wright3}. Illustration of the spectral layout, estimated spectral distributions across the Hawaii-4RG detectors, and assembled 3D data cubes are presented in Figure \ref{fig:slicer_fig}.

The Liger DRS refers to all aspects of reducing data from the Liger detectors. This includes the processes performed on the raw detector data outside the scope of the Data Reduction Pipeline (DRP) which is itself a subset of the broader DRS. While these terms are sometimes used interchangeably, although the specific difference is illustrated in Figure \ref{fig:liger_wmko} and the primary focus of this paper is discussion of the DRP as done in Section \ref{sec:liger_drp}.

Several other advanced instruments at WMKO have demonstrated the need for automated data reduction, including NIRSPEC\cite{nirspec_drp_koa, nirspec_drp_koa2} and OSIRIS\cite{osiris_drp}. In order to enable maximum versatility in science products, the reduced science data, raw unprocessed frames, and calibration data will all be made available to astronomers through the Keck Observatory Archive (KOA), as discussed in Section \ref{sec:wmko_software}.  Data products, algorithms, and other DRP characteristics are discussed in Section \ref{sec:liger_drp}.

\subsection{Heritage from IRIS}
\label{sec:iris_inheritance}
One of the fundamental advantages that Liger possesses as an instrument is design inheritance from preceding instruments, particularly IRIS, and the DRP is no exception. The DRP inherits its design directly from the IRIS DRS\cite{Walth2, Walth1, Surya1}, which itself leverages the Data Reduction System infrastructure from Space Telescope\cite{stpipe1, stpipe2}. The fundamental structure of the DRP design, like that of the IRIS Data Reduction System, is modeled after the JWST `stpipe` package which defines a framework to implement all data reduction algorithms as individual steps, combine them into pipelines, and then configure and execute them. The pipeline python package for Liger will be called 'liger\_pipeline' and mirrors the IRIS design inherited from Space Telescope. This system architecture is further discussed in Section \ref{sec:liger_drp}.

\section{Software Architecture at WMKO}

Figure \ref{fig:liger_wmko} illustrates the integration of DRS components into the Observatory software system. The readouts are assembled into full “Raw Frames” on each of the detector computers, one for the imager and one for the IFS. Each detector target computer acquires the readouts from the detector hardware and calls functions from the  Markury ASIC Control and Interface Electronics (MACIE) C library to perform low-level reduction algorithms like sampling, and then saves to disk the raw science frame for further processing. The output data files are saved in a shared high-capacity file system which resides on the pipeline computer.  The DRS (and Liger software generally) will utilize the Flexible Image Transport System (FITS)\cite{fits} format for data products which will enable easy and historically-motivated association of multiple data images via extensions and metadata via headers. Although this is not currently planned, it would not be difficult for the pipeline computer to retrieve additional telemetry or other metadata as needed and this may be incorporated as ongoing development requires.

As soon as an observation starts, the Liger Pipeline will receive a command to prepare for execution and will start to monitor the file system for raw science frame files to be available as soon as the Detector Target computers write them. When a raw science frame is available on the disk, the DRP will configure the pipeline execution based upon the available metadata and launch a `liger\_pipeline` process for each raw science frame. Therefore multiple instances of the Python pipeline will run concurrently, which will be important to enable simultaneous observation utilizing both IFS and imager modes.

The `liger\_pipeline` Python processes, once configured with the parameters for the current observing mode, will process the input raw science frames into reduced science frames in less than 15 seconds for the imager, 30 seconds for the IFS, and will also read from the local storage disk to retrieve the necessary calibration files. The suitable calibration files will be selected based on the pipeline configuration and the observation metadata retrieved from telemetry and the header of the raw frame. Once the pipelines have completed execution, they will write the output FITS files to the DRP storage disk. At this point the DRS can publish the event of "execution completed" and the data products can be made available for quicklook visualization.

The different computer systems allocated to handle the system processes illustrated in Figure \ref{fig:liger_wmko} are each catered the specific computational requirements of their respective functionalities.  For the pipeline computer, the currently budgeted specifications entail Dual Xeon Processors with 32 Gb RAM and 1+ Tb RAID-5 for data storage.  This should provide ample processing power and storage capabilities for the DRP requirements, although the exact method of transfer or shared storage space enabling writing raw science frames from the detector target computers to the pipeline computers is still to be determined as discussed in Section \ref{sec:wmko_software}.

\subsection{Software Interfaces at Observatory Level}
\label{sec:wmko_software}
Interfaces to subsystems at WMKO are managed through the Keck Task Library (KTL)\cite{ktl}, which the instrument host computer will utilize to sequence observations from astronomer-enabled scripts or commands from an observational planning graphical user interface. These sequences will be passed to the detector target computers, which will clock the detectors. Metadata associated with the particular observation sequenced from the instrument host computer will be associated with individual reads by the detector target computers after retrieval from observatory systems (also via KTL's keyword-associated metadata retrieval).

Liger software components are grouped using a logical bundle of keywords formed with one or more dispatchers. This approach is widely used amongst several successfully commissioned Keck instruments such as NIRSPEC, OSIRIS, and MOSFIRE. There are six main services provided by Liger software: two Detector Services (for the imager and spectrograph), Motion Control Service, Housekeeping Service, Power Service, and Header Service. Each service has one or more dispatchers publishing keywords to KTL, mostly current status and readings of the electronics and hardware for which each dispatcher is responsible. The exception is the Header Service, which aggregates telemetries from other dispatchers to prepare for the FITS file header. Because this paper is concerned primarily with the DRP, we will not further discuss the specifics of Liger software pertaining to each of the listed Liger services. 

\begin{figure}[h]
    \centering
    \includegraphics[scale=0.55]{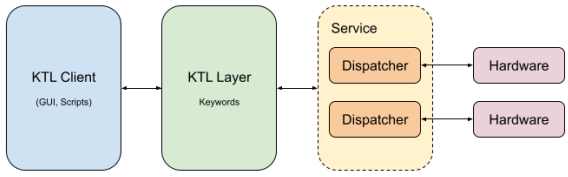}
    \caption{Basic outline of the functionality provided by KTL and how Liger will interface with subsystems. Illustration of individual subsystem interfaces is shown in Figure \ref{fig:liger_wmko}.}
    \label{fig:ktl_sketch}
\end{figure}

Besides the listed services with their respective dispatchers, several KTL client components are part of Liger software as well: Graphic User Interface (GUI), Scripts, and Data Visualization. The GUI will be the main interface to the users (astronomers and supporting staff) while scripts can help automate some of the often used, highly repeated processes to save time on observing nights.

\begin{figure}[h]
    \centering
    \includegraphics[scale=0.32]{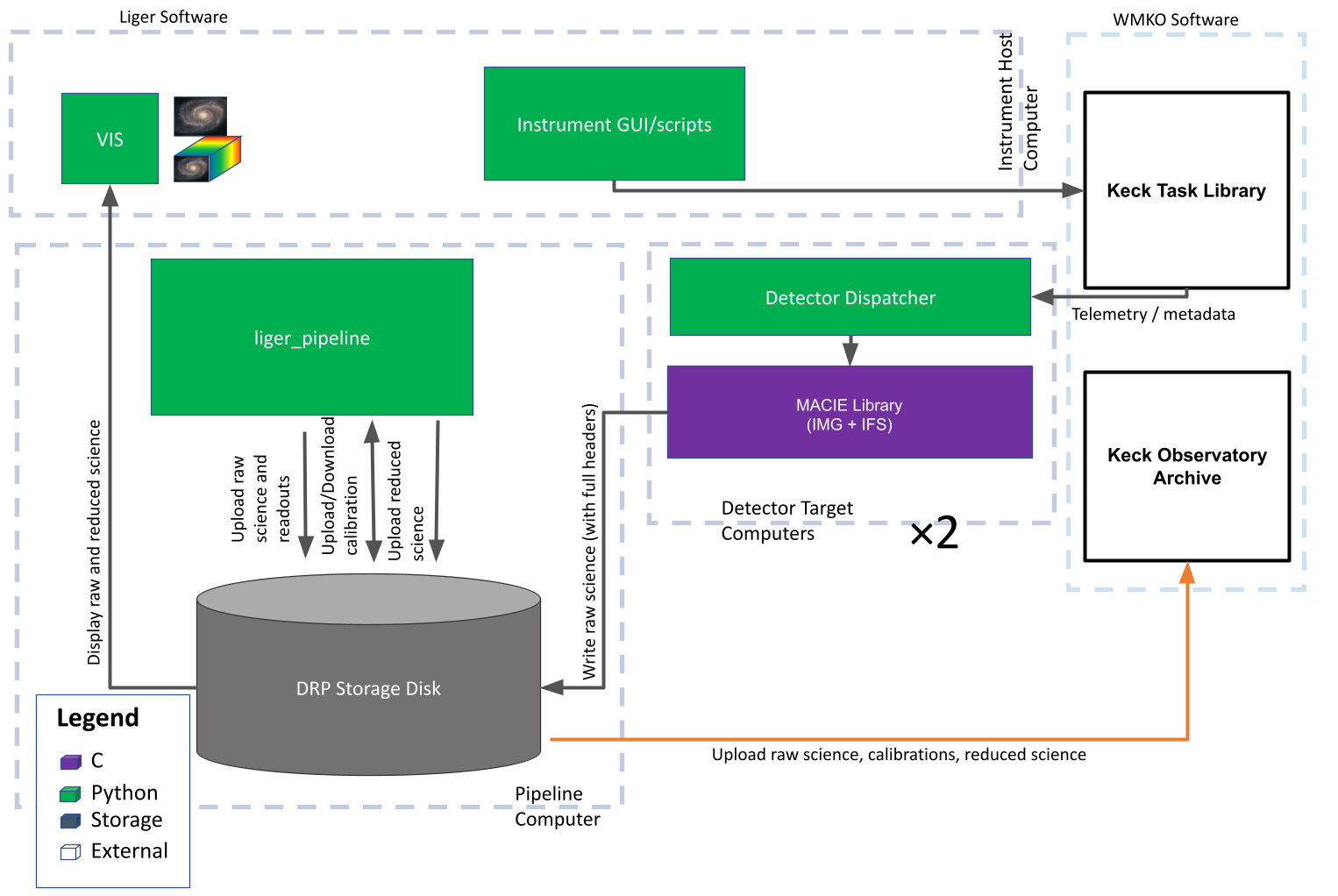}
    \caption{Liger software and WMKO software architecture and interfaces. The different computers to be used for hosting Liger software are shown in boxes, with observatory and the reduction pipelines integrated into the observatory software for online processing on site. Software packages are illustrated as the colored boxes, with color associated with coding language and uncolored boxes associated with the external software of WMKO. The rough data flow of a hypothetical observation is shown through the sequence of arrows beginning from the instrument host computer. \label{fig:liger_wmko}}
\end{figure}

Offline reduction capabilities will be provided to the astronomer via access to the KOA\cite{koa1, koa2} as illustrated in Figure \ref{fig:liger_wmko}. The format for data products and associated metadata to be ingested by KOA for astronomer access will be consistent with WMKO best practices for archival to enable clear definition and categorization\cite{koa_metadata, koa_metadata2}. Through KOA Keck astronomers will be able to download raw science frames, calibrations, and reduced data products.  The DRP code itself will be available for astronomer download and personalization via a github repository. 

\section{Data Reduction Pipeline}
\label{sec:liger_drp}

The DRP is planned to provide real-time ($<$ 1 minute) data processing of imaging and spectroscopic data, as well as a full offline reduction package. There will be three modes of the real-time data reductions on raw science frames: imager, lenslet IFS, and slicer IFS data. The visualization method for Liger data products is yet to be determined but will be compliant with WMKO software infrastructure and will take place on the instrument host computer as illustrated in Figure \ref{fig:liger_wmko}.  The Liger DRP software structure will be implemented as the raw data frames (assembled from individual ramps read from each detector target computer) are delivered to the pipeline computer from the detector target computers.

The architecture of the full DRP package will be a pipeline, the model used for many existing instruments. In software, a pipeline is a chain of processing elements (i.e. algorithms) arranged so that the output of each element is the input of the next. The DRP system will serve to link all the algorithms together and provide the necessary software infrastructure. All Liger data reduction algorithms will be custom-designed for Liger final data products. The basis of some algorithms will be adapted from previous IFS instruments, such as OSIRIS, GPI, NIFS, and SINFONI pipelines. Numerous near-infrared imagers exist (e.g. NIRC2, NACO), and they will be leveraged to provide algorithms for the Imaging mode whenever possible. The DRP will be written in Python to process data and metadata utilizing the FITS format. As discussed in Section \ref{sec:iris_inheritance}, the structure of the DRP is based on the JWST pipeline 'stpipe', the benefit of which is to provide a substantial infrastructure for defining different reduction routines and enabling modularity in software design to improve overall functionality and long-term support sustainability. 

\subsection{Motivation for STPIPE as Pipeline Architecture}

Utilizing 'stpipe' software architecture provides many benefits and increases the effectiveness of the DRP from a software design perspective.  In contrast to observatories or instruments which implement their own pipeline from scratch, Liger (Like IRIS) will adopt the general-purpose modular design infrastructure already released by Space Telescope and will not be encumbered by the need to "reinvent the wheel" in the case of data reduction pipelines by beginning from nothing.  The modularity and astronomer-specific capabilities (such as to save data products between any pipeline operation) which are well documented, supported, and tested by the Space Telescope team allow the package to be easily adapted by Liger software development and to astronomers who will wish to download the package and make alterations as necessary for their specific data reduction needs.  As more astronomers use JWST and become familiar with the pipeline associated with it, that familiarity will transfer neatly over to both IRIS and Liger data reduction systems. 

The risks involved with the adoption of 'stpipe' as a software architecture primarily lie in the use of an external software package for critical functions. Because of the extensive testing, commenting, and documentation available for 'stpipe' this risk is assessed as minimal and acceptable for the benefits of adopting this as infrastructure for the Liger DRP.

\subsection{DRP Algorithms}
\label{sec:algorithms}

The Liger pipeline will be comprised of a series of algorithms feeding the output of one as the input of the next to perform the basic data processing techniques required of Liger data products. This section will provide a description of the algorithms to be incorporated into the Liger DRP and describe the data flow. 

\begin{figure}[h!]
    \centering
    \includegraphics[scale=0.35]{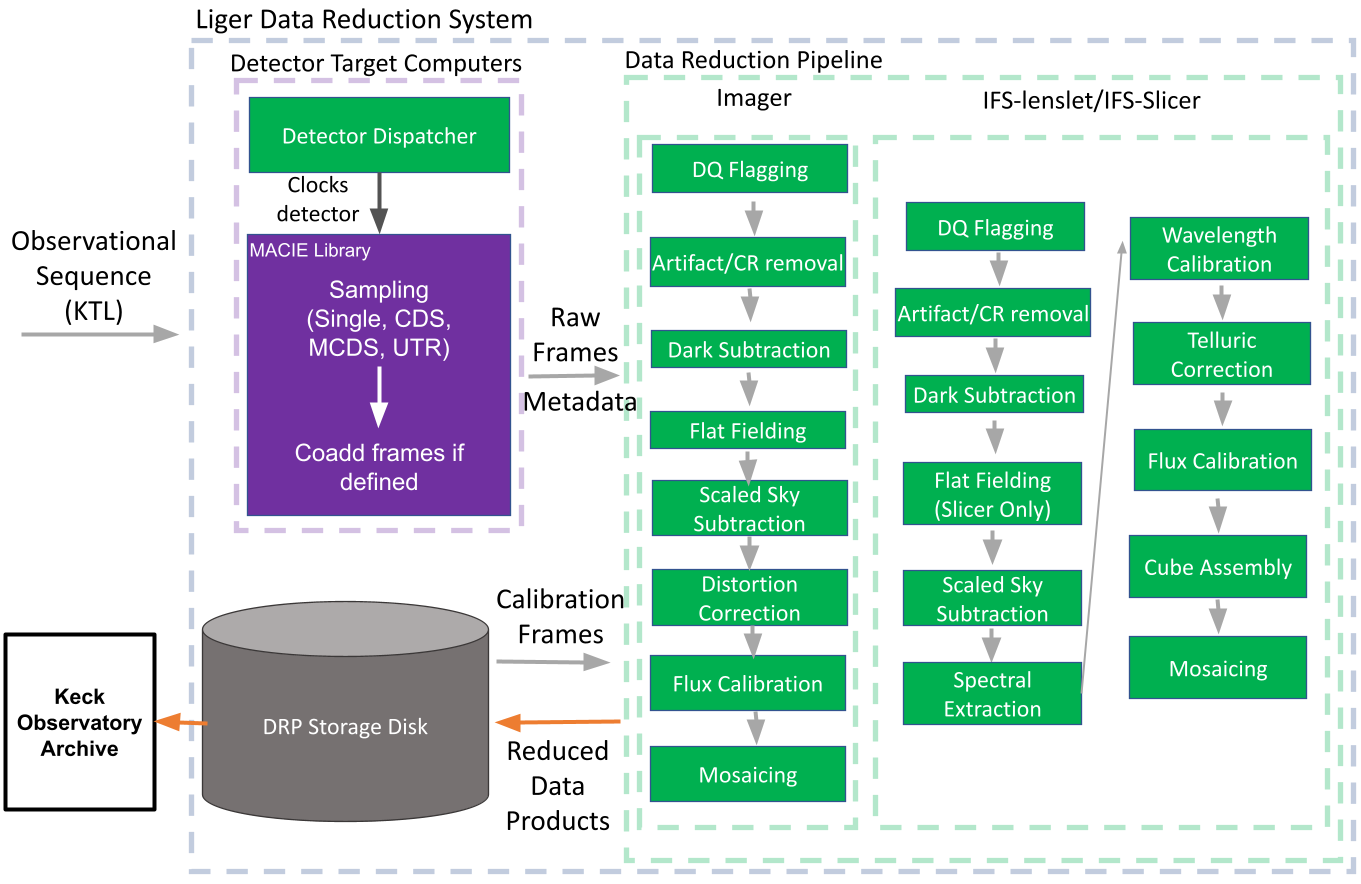}
    \caption{Data flow and processing within the Liger DRS.  IFS and Imager reduction pipelines will be capable of running concurrently (via parallelized processes) in order to reduce the processing time required for quick-look visualization on an observation night in which both the IFS and Imager are used for a single observational sequence.}
    \label{fig:drs_dataflow}
\end{figure}

As discussed in Section \ref{sec:wmko_software}, the Liger pipeline will be available for offline data processing and reduced data products will be available for visualization in real-time. However, because the processing time required for the real-time reduction for visualization upon an observation night is time-sensitive, some of the steps of the IFS and Imager pipelines are skipped during real-time reduction. Table \ref{tab:alg} gives information on algorithms that will be used in each of the offline and real-time scenarios.  There will be a separate flow for the data processing algorithms listed in Table \ref{tab:alg} depending on the selected instrument mode. In keeping with design and algorithm inheritance from IRIS DRS as discussed in Section \ref{sec:iris_inheritance}, the categorization of these algorithms mirrors the IRIS design presented in Walth et al (2018)\cite{Walth2} excluding IRIS-specific algorithms. An illustration of the data flow within the Liger DRS and the different algorithms used for each subsystem is illustrated in Figure \ref{fig:drs_dataflow}.

\begin{table}[h]
\label{tab:alg}
\begin{center}
\caption{Algorithms used for the imager and IFS real-time and final data pipelines.}
\begin{tabular}{|l|c|c|c|c|}
\hline
Algorithms & \multicolumn{2}{|c|}{Real-time Processing} & \multicolumn{2}{|c|}{Offline processing} \\
\cline{2-5} & Imager & IFS & Imager & IFS \\
\hline
Generate master dark              & x & x & x & x \\
Dark subtraction                  & x & x & x & x \\
Correction of detector artifacts  & x & x & x & x \\
Remove cosmic rays                &  &  & x & x \\
Flat fielding*                    & x & x & x & x \\
Spectral extraction               &   & x &   & x \\
Wavelength calibration            &   & x &   & x \\
Cube assembly                     &   & x &   & x \\
Scaled sky-subtraction            & x & x & x & x \\
Telluric correction               &   &   &   & x \\
Field distortion correction       &   &   & x &   \\
Flux calibration                  &   &   & x & x \\
Mosaic/Combine science                &   &   & x & x \\
\hline
\end{tabular}
\small
\item *IFS flat fielding algorithm is for slicer mode only.
\end{center}
\end{table}

The Liger DRP is comprised of the sequence of algorithms illustrated in Figure \ref{fig:drs_dataflow}, in addition to those that generate the requisite calibration frames and metadata objects to be associated with a Liger data product.  For further data analysis and processing involving more specific reduction techniques beyond those here described, the astronomer is expected to download the Liger DRP and any data products necessary for offline processing as discussed in Section \ref{sec:wmko_software}.

\subsubsection{Assign WCS}
\label{sec:wcs}
Each Liger observation will be associated with a World Coordinate System (WCS) object which will be generated from the metadata associated with each observation. Not only is this crucial metadata for the scientific viability of an observation, but this will be utilized for proper assembly of dithered frames as described in Section \ref{sec:mosaic}.
    
\subsubsection{Generate Master Dark}
\label{sec:master_dark}
Dark frames will be stored on the DRP storage disk with associated header metadata for cataloguing as calibration frames.  Upon reduction of science data these frames will be assembled into a master dark frame to be used in dark current subtraction as discussed in Section \ref{sec:dark_subt}.

\subsubsection{Data Quality Flagging}
\label{sec:dq_flag}
Each frame supplied to the DRP will be analyzed to ascertain data quality.  Data Quality (DQ) FITS extensions will be appended to each frame in order to mark bad pixels and flag other potential quality issues in the form of a bad pixel map.

\subsubsection{Detector Artifact and Bad Pixel Removal}
Utilizing the FITS extensions discussed in Section \ref{sec:dq_flag}, each frame will have detector artifacts and bad pixels removed prior to full data reduction in order to preserve the quality of science data.

\subsubsection{Cosmic Ray Removal}
Cosmic rays are unpredictable high energy photons which may be observed from the sky and cause bright artifacts to be read on the detector.  In order to preserve science data quality, each frame will be flagged for cosmic rays and the affected pixels will be removed from the frame.

\subsubsection{Flat Fielding}
Each science frame must be normalized to remove the detector variance in the field across the image.  In order to accomplish this flat field calibration frames will be taken and stored on the DRP storage disk, and normalized by dividing an input flat frame by its median.  This will then be used to scale the science and error extensions in order to remove field variance.  

\subsubsection{Dark Subtraction}
\label{sec:dark_subt}
Dark current is a consistent noise source in the form of observed current from infrared emission.  Utilizing the master dark generated as discussed in Section \ref{sec:master_dark} to reduce the effects of intrinsic photon noise, dark current will be subtracted from the detector frames.  If necessary to reduce the time impact of assembling the master dark, for real-time reduction prior to data visualization on an observation night only a single dark frame will be used to subtract dark current.

\subsubsection{Scaled Sky Subtraction}
The intrinsic brightness of the sky must be subtracted off of the science frame for optimal scientific viability.  For the imager, this can be accomplished through subtraction of a single constant value associated with the integrated brightness of the sky for that bandwidth.  For the IFS, a sky frame must be taken and scaled to match the sky brightness of the science frame.

\subsubsection{Flux Calibration}
Frames from the detector are stored with pixel values in data units, with corresponding gain values which associate the number of electrons per second per data unit.  These electrons per second are each associated with a certain number of photons per second observed from the sky.  Each frame will therefore need to be converted from data units into surface brightness units: flux density erg/s/$cm^{2}$/Hz for the imager and erg/s/$cm^{2}$/{\AA} for the IFS.  This will be accomplished through observation of a standard star calibration.

\subsubsection{Mosaic / Combine Frames}
\label{sec:mosaic}
Observational frames that have been dithered need to be combined based on their relative positions.  As discussed in Section \ref{sec:wcs}, this will be performed based on each frames WCS metadata in order to match the relative locations of specific pixels.  Both integer and non-integer numbers of pixel shifts will be supported, and either median or averages of the overlapping field regions will be allowed to combine dithered frames.

\subsubsection{Distortion Correction}
Correcting distortion of an imager frame is an essential duty of the Liger DRP. The DRP will utilize a pre-assembled distortion solution to correct each observed imager field of the instrument-specific aberrations affecting the frame. 

\subsubsection{Spectral Extraction}
\label{sec:spectral_ext}
The spectral extraction of the IFS modes of Liger is one of the most important and potentially problematic roles of the Liger DRP.  For the Slicer, this will be comprised of a relatively straightforward extraction along the locations of the spectra associated with each spatial element for the specified slicer mode.  For the lenslet, each spectra will need to be deconvolved in order to account for the spectral overlap with neighboring spectra associated with different spatial elements. The lenslet spectral extraction will require a rectification matrix. The importance and complexity of this operation is illustrated by the spectral distribution simulations shown in Figure \ref{fig:slicer_fig}.
    
\subsubsection{Wavelength Calibration}
\label{sec:wave_cal}
Proper calibration of the wavelength elements associated with each spectral element for the Liger IFS is a crucial data reduction step for the lenslet and slicer data products. This will be conducted using data frames associated with arc lamps containing typical emission lines for the near infrared such as argon, krypton, and xenon. These frames will then be used to compute the wavelength solution for each spectrum, and apply it to the extracted spectra discussed in Section \ref{sec:spectral_ext}.
    
\subsubsection{Telluric Correction}
Effects of the Earth's atmosphere are problematic for astronomical observation, and so removing the absorption features (i.e. the telluric) of the Earth's atmosphere is an important step of the Liger DRP.  This will be done through calibration from observation of a telluric A-type standard calibration star (e.g. Vega).
    
\subsubsection{Cube Assembly}
\label{sec:cubes}
Following proper spectral extraction as discussed in Section \ref{sec:spectral_ext} and calibration as discussed in Section \ref{sec:wave_cal}, IFS spectra must be assembled into 3D (X, Y, $\lambda$) data cubes to be delivered to the astronomer. The Liger DRP will assemble the extracted and calibrated spectra in accordance with their individual spatially associated elements into a data cube with WCS coordinates.

\subsection{Data Products}
The Liger DRP will deliver the following data products presented in this section to the astronomer and to the KOA for archiving. 

\subsubsection{Raw Frames}
As shown in Figure \ref{fig:liger_wmko}, the detector target computers will deliver all raw frames to the Liger DRP.  The processing algorithms described in Section \ref{sec:algorithms} will take place, but the unprocessed 2D raw detector frames will also be passed through the DRP to enable any offline processing the astronomer desires. 

\subsubsection{2D Reduced Frames}
The DRP will reduce the input frames from the detector target computers as described in Section \ref{sec:algorithms}.  The 2D processed imager and calibration frames will then be passed to KOA for science frame availability and for potential offline additional processing by astronomer.

\subsubsection{3D Data Cubes}
IFS processed data products take the form of 3D (X, Y, $\lambda$) data cubes as discussed in Section \ref{sec:cubes}. These are then delivered by the DRP. 

\subsubsection{DQ Extensions}
All detector frames will have associated data quality extensions, as discussed in Section \ref{sec:dq_flag}. These will be delivered with all FITS data products to illustrate potential quality issues with associated detector pixels. 

\subsubsection{Metadata} 
As illustrated by Figures \ref{fig:liger_wmko} and \ref{fig:drs_dataflow} and discussed in Section \ref{sec:algorithms}, metadata will be used by the DRP to conduct many necessary processes and will be associated with all Liger DRP data products in the form of FITS headers. Telemetry, observatory, and instrument metadata will be delivered with every Liger data product to enable optimal scientific viability for the groundbreaking astronomy that Liger will undoubtedly allow.

\section{Summary}
Liger is an exciting new instrument for WMKO that is designed with heritage from IRIS and OSIRIS to fully take advantage of the new KAPA AO system.  As both an imager and integral field spectrograph with hundreds of potential modes, a robust data reduction pipeline is a necessary addition to the instrument software infrastructure.  We have presented the initial design for the Liger DRS, utilizing stpipe as architecture and incorporating the DRS into WMKO software.  Development of the DRS is ongoing, and future iterations of its design will further streamline the software interfaces and algorithms involved to ensure that optimally science-ready data products be delivered upon Liger's first light.

\acknowledgments     
This research program has been made possible by the Heising-Simons Foundation (Grant \#2018-1085: Wright) and the Gordon and Betty Moore Foundation (Grant \#11169: Wright).


\bibliographystyle{spiebib}
\bibliography{report.bib}   

\begin{thebibliography}{10}

\bibitem{Larkin1}
Larkin, J.~E. and et~al, ``{The Infrared Imaging Spectrograph (IRIS) for TMT:
  instrument overview},'' in [{\em Ground-based and Airborne Instrumentation
  for Astronomy VII}{\nolinebreak\hspace{0.1em}]},  SPIE (July 2018).

\bibitem{Larkin2}
Larkin, J.~E., Moore, A.~M., Wright, S.~A., Wincentsen, J.~E., Anderson, D.,
  Chisholm, E.~M., Dekany, R.~G., Dunn, J.~S., Ellerbroek, B.~L., Hayano, Y.,
  Phillips, A.~C., Simard, L., Smith, R., Suzuki, R., Weber, R.~W., Weiss,
  J.~L., and Zhang, K., ``{The Infrared Imaging Spectrograph (IRIS) for TMT:
  instrument overview},'' in [{\em Ground-based and Airborne Instrumentation
  for Astronomy VI}{\nolinebreak\hspace{0.1em}]},  SPIE (August 2016).

\bibitem{Larkin3}
Moore, A.~M., Larkin, J.~E., Wright, S.~A., Bauman, B., Dunn, J., Ellerbroek,
  B., Phillips, A.~C., Simard, L., Suzuki, R., Zhang, K., Aliado, T., Brims,
  G., Canfield, J., Chen, S., Dekany, R., Delacroix, A., Do, T., Herriot, G.,
  Ikenoue, B., Johnson, C., Meyer, E., Obuchi, Y., Pazder, J., Reshetov, V.,
  Riddle, R., Saito, S., Smith, R., Sohn, J.~M., Uraguchi, F., Usuda, T., Wang,
  E., Wang, L., Weiss, J., and Wooff, R., ``{The Infrared Imaging Spectrograph
  (IRIS) for TMT: instrument overview},'' in [{\em Ground-based and Airborne
  Instrumentation for Astronomy V}{\nolinebreak\hspace{0.1em}]},  SPIE (August
  2014).

\bibitem{Larkin4}
Larkin, J.~E., Moore, A.~M., Barton, E.~J., Bauman, B., Bui, K., Canfield, J.,
  Crampton, D., Delacroix, A., Fletcher, M., Hale, D., Loop, D., Niehaus, C.,
  Phillips, A.~C., Reshetov, V., Simard, L., Smith, R., Suzuki, R., Usuda, T.,
  and Wright, S.~A., ``{The Infrared Imaging Spectrograph (IRIS) for TMT:
  instrument overview},'' in [{\em Ground-based and Airborne Instrumentation
  for Astronomy III}{\nolinebreak\hspace{0.1em}]},  SPIE (July 2010).

\bibitem{Wright2020}
{Wiley}, J., {Mathur}, K., {Brown}, A., {Wright}, S., {Cosens}, M., {Maire},
  J., {Fitzgerald}, M., {Jones}, T., {Kassis}, M., {Kress}, E., {Kupke}, R.,
  {Larkin}, J.~E., {Lyke}, J., {Wang}, E., and {Yeh}, S., ``{Liger for
  next-generation Keck adaptive optics: opto-mechanical dewar for imaging
  camera and slicer},'' in [{\em Society of Photo-Optical Instrumentation
  Engineers (SPIE) Conference Series}{\nolinebreak\hspace{0.1em}]},  {\em
  Society of Photo-Optical Instrumentation Engineers (SPIE) Conference Series}
  {\bf 11447},  1144758 (Dec. 2020).

\bibitem{Wright2022}
Wright, S.~A., ``{Liger at Keck Observatory: Overall Design Specifications and
  Science Drivers},'' in [{\em Ground-based and Airborne Instrumentation for
  Astronomy IX}{\nolinebreak\hspace{0.1em}]},  SPIE (July 2022).

\bibitem{Wizinowich}
Wizinowich, P., Lu, J., and Cetre, S., ``{Keck All Sky Precision Adaptive
  Optics Program Overview},'' in [{\em Ground-based and Airborne
  Instrumentation for Astronomy IX}{\nolinebreak\hspace{0.1em}]},  SPIE (July
  2022).

\bibitem{Cosens2020}
{Cosens}, M., {Wright}, S.~A., {Arriaga}, P., {Brown}, A., {Fitzgerald}, M.,
  {Jones}, T., {Kassis}, M., {Kress}, E., {Kupke}, R., {Larkin}, J.~E., {Lyke},
  J., {Wang}, E., {Wiley}, J., and {Yeh}, S., ``{Liger for next-generation Keck
  AO: filter wheel and pupil design},'' in [{\em Ground-based and Airborne
  Instrumentation for Astronomy VIII}{\nolinebreak\hspace{0.1em}]},  {\em
  Society of Photo-Optical Instrumentation Engineers (SPIE) Conference Series}
  {\bf 11447},  114474X (Dec. 2020).

\bibitem{Cosens2022}
{Cosens}, M., {Wright}, S.~A., {Brown}, A., {Fitzgerald}, M., {Johnson}, C.,
  {Jones}, T., {Kassis}, M., {Kress}, E., {Kupke}, R., {Larkin}, J.~E.,
  {Magnone}, K., {McGurk}, R., {Rundquist}, N.-E., {Sohn}, J.~M., {Wang}, E.,
  {Wiley}, J., and {Yeh}, S., ``{Liger at Keck Observatory: Imager Detector and
  IFS Pick-off Mirror Assembly},'' in [{\em Ground-based and Airborne
  Instrumentation for Astronomy IX}{\nolinebreak\hspace{0.1em}]},  {\em Society
  of Photo-Optical Instrumentation Engineers (SPIE) Conference Series}, SPIE
  (July 2022).

\bibitem{Whiley}
Whiley, J., ``{Liger at Keck Observatory: Final Design of Imager Cryogenic
  Dewar and Spectrograph Re-Imaging Optic},'' in [{\em Ground-based and
  Airborne Instrumentation for Astronomy IX}{\nolinebreak\hspace{0.1em}]},
  SPIE (July 2022).

\bibitem{Wright1}
Wright, S.~A., Walth, G., Do, T., Marshall, D., Larkin, J.~E., Moore, A.~M.,
  Adamkovics, M., Andersen, D., Armus, L., Barth, A., Cote, P., Cooke, J.,
  Chisholm, E.~M., Davidge, T., Dunn, J.~S., Dumas, C., Ellerbroek, B.~L.,
  Ghez, A.~M., Hao;, L., Hayano, Y., Liu, M., Lopez-Rodriguez, E., Lu;, J.~R.,
  Mao, S., Marois, C., Pandey, S.~B., Phillips, A.~C., Schoeck, M.,
  Subramaniam, A., Subramanian, S., Suzuki, R., Tan, J.~C., Terai, T., Treu,
  T., Simard, L., Weiss, J.~L., Wincentsen, J., Wong, M., and Zhang, K., ``{The
  infrared imaging spectrograph (IRIS) for TMT: latest science cases and
  simulations},'' in [{\em Adaptive Optics Systems
  V}{\nolinebreak\hspace{0.1em}]},  SPIE (July 2016).

\bibitem{Wright2}
Wright, S.~A., Larkin, J.~E., Moore, A.~M., Do, T., Simard, L., Adamkovics, M.,
  Armus, L., Barth, A.~J., Barton, E., Boyce, H., Cooke, J., Cote, P., Davidge,
  T., Ellerbroek, B., Ghez, A.~M., Liu, M.~C., Lu, J.~R., Macintosh, B.~A.,
  Mao, S., Marois, C., Schoeck, M., Suzuki, R., Tan, J.~C., Treu, T., Wang, L.,
  and Weiss, J., ``{The infrared imaging spectrograph (IRIS) for TMT: overview
  of innovative science programs},'' in [{\em Ground-based and Airborne
  Instrumentation for Astronomy V}{\nolinebreak\hspace{0.1em}]},  SPIE (July
  2014).

\bibitem{Wright3}
Wright, S.~A., Barton, E.~J., Larkin, J.~E., Moore, A.~M., Crampton, D., and
  Simard, L., ``{The infrared imaging spectrograph (IRIS) for TMT:
  sensitivities and simulations},'' in [{\em Ground-based and Airborne
  Instrumentation for Astronomy III}{\nolinebreak\hspace{0.1em}]},  SPIE (July
  2010).

\bibitem{nirspec_drp_koa}
{Tran}, H.~D., {Cohen}, R., {Mader}, J.~A., {Colson}, A., {Berriman}, G.~B.,
  {Gelino}, C.~R., and {KOA Team}, ``{Data reduction pipelines for the Keck
  Observatory Archive},'' in [{\em American Astronomical Society Meeting
  Abstracts \#227}{\nolinebreak\hspace{0.1em}]},  {\em American Astronomical
  Society Meeting Abstracts} {\bf 227},  348.23 (Jan. 2016).

\bibitem{nirspec_drp_koa2}
{Mader}, J.~A., {Tran}, H.~D., {Cohen}, R., {Colson}, A., {Berriman}, G.~B.,
  {Gelino}, C.~R., {Kong}, M., {Laity}, A.~C., {Swain}, M.~A., {Wang}, C.,
  {Goodrich}, R., and {Holt}, J., ``{The Design and Development of the NIRSPEC
  Data Reduction Pipeline for the Keck Observatory Archive},'' in [{\em
  Astronomical Data Analysis Software and Systems
  XXV}{\nolinebreak\hspace{0.1em}]},  {Lorente}, N.~P.~F., {Shortridge}, K.,
  and {Wayth}, R., eds., {\em Astronomical Society of the Pacific Conference
  Series} {\bf 512},  399 (Dec. 2017).

\bibitem{osiris_drp}
{Lockhart}, K.~E., {Do}, T., {Larkin}, J.~E., {Boehle}, A., {Campbell}, R.~D.,
  {Chappell}, S., {Chu}, D., {Ciurlo}, A., {Cosens}, M., {Fitzgerald}, M.~P.,
  {Ghez}, A., {Lu}, J.~R., {Lyke}, J.~E., {Mieda}, E., {Rudy}, A.~R., {Vayner},
  A., {Walth}, G., and {Wright}, S.~A., ``{Characterizing and Improving the
  Data Reduction Pipeline for the Keck OSIRIS Integral Field Spectrograph},''
  {\em The Astronomical Journal}~{\bf 157},  75 (Feb. 2019).

\bibitem{Walth2}
Walth, G., Wright, S.~A., Weiss, J., Larkin, J.~E., Moore, A.~M., Chapin,
  E.~L., Do, T., Dunn, J., Ellerbroek, B., Gillies, K., Hayano, Y., Johnson,
  C., Marshall, D., Riddle, R.~L., Simard, L., Sohn, J.~M., Suzuki, R., and
  Wincentsen, J., ``{The Infrared Imaging Spectrograph (IRIS) for TMT: data
  reduction system},'' in [{\em Software and Cyberinfrastructure for Astronomy
  IV}{\nolinebreak\hspace{0.1em}]},  SPIE (August 2016).

\bibitem{Walth1}
Walth, G., Wright, S.~A., Weiss, J., Larkin, J.~E., Moore, A.~M., Chapin,
  E.~L., Do, T., Dunn, J., Ellerbroek, B., Gillies, K., Hayano, Y., Johnson,
  C., Marshall, D., Riddle, R.~L., Simard, L., Sohn, J.~M., Suzuki, R., and
  Wincentsen, J., ``{The Infrared Imaging Spectrograph (IRIS) for TMT:
  advancing the DRS},'' in [{\em Software and Cyberinfrastructure for Astronomy
  V}{\nolinebreak\hspace{0.1em}]},  SPIE (July 2018).

\bibitem{Surya1}
Surya, A., Zonca, A., Rundquist, N., Wright, S.~A., Walth, G., Weiss, J.,
  Larkin, J.~E., Moore, A.~M., Chapin, E.~L., Do, T., Dunn, J., Ellerbroek, B.,
  Gillies, K., Hayano, Y., Johnson, C., Marshall, D., Riddle, R.~L., Simard,
  L., Sohn, J.~M., Suzuki, R., and Wincentsen, J., ``{The Infrared Imaging
  Spectrograph (IRIS) for TMT: Final Design Development of the Data Reduction
  System},'' in [{\em Software and Cyberinfrastructure for Astronomy
  VI}{\nolinebreak\hspace{0.1em}]},  SPIE (July 2020).

\bibitem{stpipe1}
{Bushouse}, H., {Droettboom}, M., and {Greenfield}, P., ``{The JWST Data
  Calibration Pipeline},'' in [{\em Astronomical Data Analysis Software and
  Systems XXV}{\nolinebreak\hspace{0.1em}]},  {Lorente}, N.~P.~F.,
  {Shortridge}, K., and {Wayth}, R., eds., {\em Astronomical Society of the
  Pacific Conference Series} {\bf 512},  355 (Dec. 2017).

\bibitem{stpipe2}
{Bushouse}, H., {Eisenhamer}, J., and {Davies}, J., ``{The JWST Data
  Calibration Pipeline},'' in [{\em Astronomical Data Analysis Software and
  Systems XXVII}{\nolinebreak\hspace{0.1em}]},  {Teuben}, P.~J., {Pound},
  M.~W., {Thomas}, B.~A., and {Warner}, E.~M., eds., {\em Astronomical Society
  of the Pacific Conference Series} {\bf 523},  543 (Oct. 2019).

\bibitem{fits}
{Grosb{\o}l}, P., ``{The FITS data format.},'' in [{\em Databases and On-line
  Data in Astronomy}{\nolinebreak\hspace{0.1em}]},  {Albrecht}, M.~A. and
  {Egret}, D., eds., {\em Astrophysics and Space Science Library} {\bf 171},
  253--257 (Jan. 1991).

\bibitem{ktl}
{Lupton}, W.~F., ``{Keck Telescope Control System},'' in [{\em Astronomical
  Data Analysis Software and Systems IX}{\nolinebreak\hspace{0.1em}]},
  {Manset}, N., {Veillet}, C., and {Crabtree}, D., eds., {\em Astronomical
  Society of the Pacific Conference Series} {\bf 216},  261 (Jan. 2000).

\bibitem{koa1}
{Berriman}, G.~B., {Gelino}, C.~R., {Laity}, A., {Kong}, M., {Swain}, M.,
  {Holt}, J., {Goodrich}, R., {Mader}, J., and {Tran}, H.~D., ``{The Operation
  and Architecture of the Keck Observatory Archive},'' in [{\em Astronomical
  Data Analysis Software and Systems XXIII}{\nolinebreak\hspace{0.1em}]},
  {Manset}, N. and {Forshay}, P., eds., {\em Astronomical Society of the
  Pacific Conference Series} {\bf 485},  123 (May 2014).

\bibitem{koa2}
{Berriman}, G.~B., {Gelino}, C.~R., {Goodrich}, R.~W., {Holt}, J., {Kong}, M.,
  {Laity}, A.~C., {Mader}, J.~A., {Swain}, M., and {Tran}, H.~D., ``{The design
  and operation of the Keck Observatory archive},'' in [{\em Software and
  Cyberinfrastructure for Astronomy III}{\nolinebreak\hspace{0.1em}]},
  {Chiozzi}, G. and {Radziwill}, N.~M., eds., {\em Society of Photo-Optical
  Instrumentation Engineers (SPIE) Conference Series} {\bf 9152},  91520A (July
  2014).

\bibitem{koa_metadata}
{Tran}, H.~D., {Holt}, J., {Goodrich}, R.~W., {Mader}, J.~A., {Swain}, M.,
  {Laity}, A.~C., {Kong}, M., {Gelino}, C.~R., and {Berriman}, G.~B.,
  ``{Metadata and data management for the Keck Observatory Archive},'' in [{\em
  Software and Cyberinfrastructure for Astronomy
  III}{\nolinebreak\hspace{0.1em}]},  {Chiozzi}, G. and {Radziwill}, N.~M.,
  eds., {\em Society of Photo-Optical Instrumentation Engineers (SPIE)
  Conference Series} {\bf 9152},  91522I (July 2014).

\bibitem{koa_metadata2}
{Berriman}, G.~B., {Holt}, J.~M., {Mader}, J.~A., {Tran}, H.~D., {Goodrich},
  R.~W., {Gelino}, C.~R., {Laity}, A.~C., {Kong}, M., and {Swain}, M.~A.,
  ``{Data and Metadata Management at the Keck Observatory Archive},'' in [{\em
  Astronomical Data Analysis Software an Systems XXIV (ADASS
  XXIV)}{\nolinebreak\hspace{0.1em}]},  {Taylor}, A.~R. and {Rosolowsky}, E.,
  eds., {\em Astronomical Society of the Pacific Conference Series} {\bf 495},
  535 (Sept. 2015).

\end{thebibliography}

\end{document}